\documentclass{article}
\usepackage{spconf}
\usepackage{enumerate}
\usepackage{cite}
\usepackage[pdftex]{graphicx}
\usepackage{amsmath}
\usepackage{array}
\usepackage{amssymb}
\usepackage{bm}
\usepackage{makecell}
\usepackage{multirow}
\usepackage{color}

\usepackage{algorithmic}
\usepackage{algorithm}

\newcommand{\eref}[1]{Eq.~(\ref{#1})}
\newcommand{\fref}[1]{Fig.~\ref{#1}}

\newcommand{\tref}[1]{Tab.~\ref{#1}}

\newtheorem{myTheo}{Theorem}

\title{Single-photon Image Super-resolution via Self-supervised Learning}
\name{Yiwei Chen, Chen Jiang, Yu Pan$^*$\thanks{This research was accepted by ICASSP 2023. Codes will be made available at https://github.com/ewellchen/SPISR-SS.}}
\address{College of Control Science and Engineering\\ Zhejiang University\\Hangzhou, China}

\begin{document}
%
\maketitle
\begin{abstract}
Single-Photon Image Super-Resolution (SPISR) aims to recover a high-resolution volumetric photon counting cube from a noisy low-resolution one by computational imaging algorithms. In real-world scenarios, pairs of training samples are often expensive or impossible to obtain. By extending Equivariant Imaging (EI) to volumetric single-photon data, we propose a self-supervised learning framework for the SPISR task. Particularly, using the Poisson unbiased Kullback-Leibler risk estimator and equivariance, our method is able to learn from noisy measurements without ground truths. Comprehensive experiments on simulated and real-world dataset demonstrate that the proposed method achieves comparable performance with supervised learning and outperforms interpolation-based methods.
\end{abstract}
\section{Introduction}
\label{sec:intro}

With rapid developments of high-end hardware and computational algorithms, single-photon imaging enables a variety of attractive applications, such as long-range imaging \cite{mccarthy2013kilometer, li2021single}, underwater imaging \cite{maccarone2019three}.  However, inevitable physical properties, e.g., the diffraction limit and minimum time resolution of constructed instruments, limit the lateral and longitudinal resolution of a single-photon imaging system. Developing effective reconstruction algorithms is a promising way to further improve the overall resolution. 

In single-photon imaging, Time-of-Fight (ToF) measurements are counted into a Photon Counting Histogram (PCH) pixel by pixel based on the Time-Correlated Single Photon Counting (TCSPC) technique \cite{becker2005advanced}. After scanning, a volumetric Photon Counting Cube (PCC) denoted by $H \in \mathbb{Z}^{T\times S\times S}$, where $S$ and $T$ are lateral and longitudinal resolution, respectively, is created by integrating all PCHs properly. In practice, $S$ is restricted by the scanning interval, which have to be larger than the Field of View (FoV) limited by the Airy disk diameter to keep all pixels from overlapping. Lateral SPISR is to overcome the FoV limit, especially for the challenging scenarios such as long-range and small-scale imaging \cite{li2020super}. Meanwhile, in regard of reducing data acquisition time \cite{rapp2017few, lindell2018single, yao2022robust}, lateral SPISR is also a viable way to reduce scanning points exponentially. Besides, the joint resolution of constructed instruments imposes a limitation on $T$, which is influenced by physical constrains that are technically challenging to enhance, such as time jitters. Longitudinal SPISR aims to shorten the minimum length of time bins in all PCHs, and thus improve the upper bound of the longitudinal resolution. Although the pileup effect distorts the waveform, shortening the longitudinal resolution is still beneficial for the imaging systems that can alleviate such effect \cite{rehain2020noise, gupta2019asynchronous}. 

The goal of SPISR is to recover a High-Resolution (HR) PCC $H^{h} \in \mathbb{R}^{T^{h}\times S^{h}\times S^{h}}$ by only a raw Low-Resolution (LR) one $H^{l} \in \mathbb{R}^{T^{l}\times S^{l}\times S^{l}}$, which inverts the forward process
\begin{equation}
	H^{l}= \mathcal{A}_{\downarrow}(H^{h} ) + \epsilon,
	\label{eq:inverse-problem}
\end{equation}
where $\mathcal{A}_{\downarrow}: \mathbb{R}^{T_{h}\times S_{h}\times S_{h}} \rightarrow \mathbb{R}^{T_{l} \times S_{l}\times S_{l}}$ is the down-sampling operator and $\epsilon$ is the independent noise term. SPISR is a challenging problem, since $H^{l}$ is corrupted by Poisson noise in practice (see \eref{eq:forward}) and the solution is seriously ill-posed \cite{kabanikhin2011inverse}. Deep Neural Network (DNNs) have achieved superior performance in SR tasks due to the scene-specific patterns learned from training data. By simplifying SPISR into depth image SR, a customized DNN showcases the ability to improve the lateral resolution of single-photon cameras guided by HR intensity images \cite{ruget2021robust}, which outperforms the interpolation-based methods \cite{keys1981cubic}. However, obtaining intensity and depth images in real-world scenarios presents significant challenges, and may even be unfeasible. 

In this work, we propose a self-supervised learning framework for SPISR. Our method is an extension of robust EI \cite{chen2022robust} to volumetric single-photon data. Specifically, we use the Poisson Unbiased Kullback-Leibler (PUKL) risk estimator \cite{bigot2017generalized} and the equivariance to tailor unsupervised losses for this task. By exploiting the rotation and shift invariance, our method is able to extract robust patterns even from a limited set of training samples. The main contributions are as follows:
\begin{itemize}
\item We propose a self-supervised learning framework to solve SPISR problem by incorporating the PUKL risk estimator and the equivariance in single-photon signals into the learning process.

\item We evaluate our model on simulated and real-world datasets, which demonstrates that it achieves comparable performance as supervised learning and superior performance to interpolation-based methods.
\end{itemize}
The paper is organized as follows. In Section \ref{sec:preliminary}, we introduce the preliminary. In Section \ref{sec:method}, we present the self-supervised framework for the SPISR problem. In Section \ref{sec:results}, the proposed method are evaluated on simulated and real-world datasets. Conclusions are given in Section \ref{sec:conclusion}.

\section{Preliminary}
\label{sec:preliminary}
\subsection{The forward model for single-photon imaging }
The photon count is modelled by a linear mixture of signal and background photons corrupted by Poisson noise. Given the system impulse response function $s(t)$ and depth $D \in \mathbb{R}^{S^{l} \times S^{l}}$, the photon count for the ${(i, j)}$-th pixel at time $t$ is
\begin{equation}
	H[i,j,t]=K \cdot \mathcal{P}\left(\eta[i,j] s(t-\frac{2D[i,j]}{\mathrm{c}})+n[i,j]\right),
	\label{eq:forward}
\end{equation}
where $\eta[i,j] \in(0,1]$ is the scale factor determined by the albedo and quantum efficiency of the detector, $n[i,j]$ is the detected noise, $K$ is the number of emitted illuminations, $\mathcal{P}(\cdot)$ denotes the Poisson process and $\mathrm{c}$ is the light speed.

\subsection{Poisson Unbiased Kullback-Leibler (PUKL) risk estimator}
For signals that are corrupted by Poisson noise, it is possible to unbiasedly estimate the Kullback-Leibler (KL) divergence by the PUKL risk estimator as follows.
\begin{myTheo}
	\label{T:PUKL} \cite{bigot2017generalized}
	Let $f: \mathbb{Z}^{n} \rightarrow \mathbb{R}^{n}$ be a measurable mapping. Let $X \in \mathbb{R}^{n}$ be the clean signal and $1 \leq i \leq n$. Let $Y \in \mathbb{Z}^{n}$ be a vector whose entries are independently sampled from a Poisson distribution of $X$. Then, 
	\begin{equation}
		\operatorname{PUKL}(f(Y),Y)=-\sum_{i=1}^n Y[i] \log \left(f(Y-e_i)[i]\right),
	\end{equation}
	where $e_i \in \mathbb{Z}^{n}$ is the $i^{th}$ canonical vector, is an unbiased estimator of $\operatorname{KL}(f(Y),X) - \sum_{i}^{T} X[i]\log X[i]$. 
\end{myTheo}

\subsection{Equivariance in deep learning}
Equivariance is the property that applying a symmetry transformation and then computing a function produces the same result as changing the order of these two operations, which has been applied to self-supervised learning \cite{chen2021equivariant, chen2022robust}, contrastive learning \cite{dangovski2022equivariant} and neural network design \cite{worrall2017harmonic}. Particularly, \cite{chen2021equivariant} proposed a self-supervised learning framework named Equivariant Imaging (EI) that efficiently exploits the natural equivariance in 2D images to solve inverse problems. To be more specific, the model learns a reconstruction function $f_\theta$ that maps the raw measurement $y$ to the signal $x$. The core assumption of EI is that the composition $f_\theta \circ \mathcal{A}$ is equivariant to a group of transformations (e.g., shifts, rotations, etc.) denoted by $\mathrm{G}=\{\mathcal{G}_1, \ldots, \mathcal{G}_{|\mathrm{G}|}\}$, formulated as
\begin{equation}
	\mathcal{G}_i(f_\theta(\mathcal{A}(x)))=f_\theta\left(\mathcal{A}\left(\mathcal{G}_i(x)\right)\right).
\end{equation}
In addition, the image $x \in \mathrm{X}$ is invariant to these transformations, that is,  $\mathcal{G}_i(x) \in \mathrm{X}$, yielding
\begin{equation}
	y=\mathcal{A}(x)=\mathcal{A}(\mathcal{G}_i(\mathcal{G}_i^{-1}(x)))=\mathcal{A}_{\mathcal{G}_{i}}(\tilde{x}), 
\end{equation}
where $\mathcal{A}_{\mathcal{G}_{i}} = \mathcal{A}\circ \mathcal{G}_{i}$ and $\tilde{x}=\mathcal{G}^{-1}_{i}(x) \in \mathrm{X}$. In this sense, transformations is equal to generate multiple virtual forward operators $\{A_{\mathcal{G}_{i}}\}_{i=1,\dots,|\mathrm{G}|}$, which make it possible to learn beyond the range space of $\mathcal{A}$. Furthermore, \cite{chen2022robust} adopted Stein’s unbiased risk estimator to improve the robustness against to noise. Following this line, we adopt the PUKL risk estimator to improve the robustness for the SPISR problem since single-photon data are corrupted by Poisson noise.

\begin{figure}
	\centering
	\includegraphics[width=0.45\textwidth]{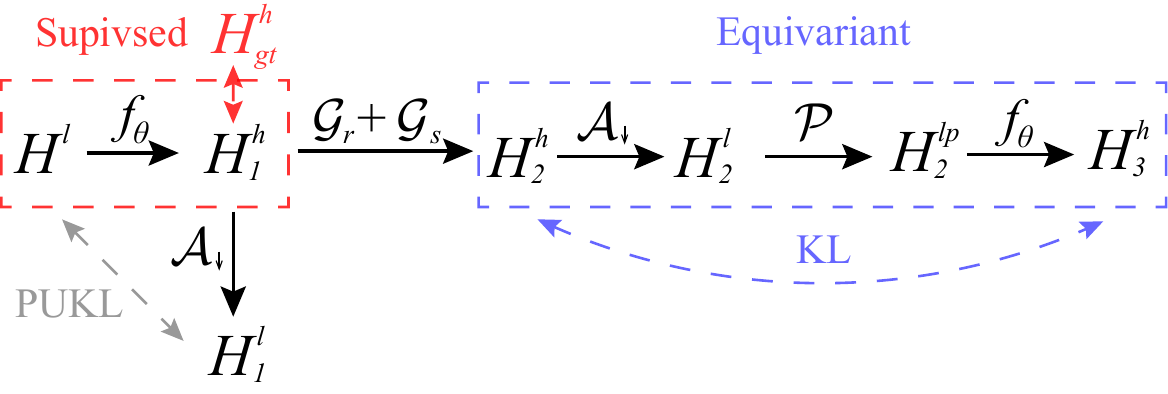}
	\caption{The pipeline of our learning framework. The supervised learning and equivariant learning process are marked by the red and blue box, respectively. The PUKL and KL losses are denoted by grey and blue dashed arrows, respectively.}
	\vspace{-10pt}
	\label{fig:re}
\end{figure}

\section{Method}
\label{sec:method}
The pipeline of our framework is shown in \fref{fig:re}. A learnable neural network defined by $f_\theta: \mathbb{R}^{T_{l}\times S_{l}\times S_{l}} \rightarrow \mathbb{R}^{T_{h} \times S_{h}\times S_{h}}$ is the solution of the SPISR problem. In supervised setting, based on the KL divergence, which is popular in the optimization of single-photon imaging models \cite{lindell2018single}, the supervised loss for SPISR is defined to learn the consistency as
\begin{equation}
	\mathcal{L}_{\mathrm{C}} = \sum_{i,j,k}^{T_{h},S_{h},S_{h}} H^{h}_{gt}[i,j,k]\log \frac{H^{h}_{gt}[i,j,k]}{H^{h}_{1}[i,j,k]},
	\label{Eq:skl}
\end{equation}
where $H^{h}_{gt} \in \mathbb{R}^{T^{h}\times S^{h}\times S^{h}}$ is the normalized ground-truth PCC and $H^{h}_{1} =  f_\theta(H^{l}) \in \mathbb{R}^{T^{h}\times S^{h}\times S^{h}}$ is the predicted PCC.
In self-supervised setting, we are only accessible to a dataset of raw LR samples $\{ [H^{l}]_{i} \}_{i = 1, \dots, N}$. The PUKL risk estimator is adopted to bypass the need of $H^{l}_{gt}$.

\subsection{Consistency learning based on PUKL}
The forward model for $(j,k)$-th pixel can be simplified as $H[j,k] = \gamma Z[j,k]$ with $Z[j,k] \sim \mathcal{P}(H_{gt}[j,k]/\gamma)$, where $\gamma$ is the parameter that controls the effect of Poisson process. Since the term $\sum_{i}^{T} H_{gt}[i,j,k]\log H_{gt}[i,j,k]$ is always a constant in this task, we define the PUKL-based loss based on Theorem~\ref{T:PUKL} as
\begin{align}
	&\mathcal{L}_{\operatorname{PUKL}}=- \sum_{i,j,k}^{T_{l},S_{l},S_{l}} H^{l}[i,j,k] \log \left(f_\theta(H^{l}-\gamma E_{ijk})\right)[i,j,k] \nonumber\\
	&\approx - \sum_{i,j,k}^{T_{l},S_{l},S_{l}} H^{l}[i,j,k] \log \left(f_\theta(H^{l})-\gamma \delta f_\theta(H^{l})\right)[i,j,k],
\end{align}
where $E_{ijk}$ is the canonical vector and we use a Taylor expansion as $f_\theta(H^{l}-\gamma E_{ijk}) \approx f_\theta(H^{l})-\gamma \delta f_\theta(H^{l})$. The Monte Carlo estimate in \cite{chen2022robust} is applied to the differential term and we get the consistency loss as
\begin{equation}
	\mathcal{L}_{\operatorname{PUKL}} = - \sum_{i,j,k}^{T_{l},S_{l},S_{l}} H^{l}[i,j,k] \log  H^{l}_{m}[i,j,k]
\end{equation}
 with
 \begin{equation}
 	H^{l}_{m} = H^{l}_1 + \frac{\gamma}{\tau} B \odot H^{l}  \odot (\mathcal{A}_{\downarrow}(f_\theta(H^{l} + \tau B))-H^{l}_1),
 \end{equation}
where $\tau$ is a small positive value, $\odot$ is the element-wise multiplication and the elements of $B \in \mathbb{R}^{T_{l} \times S_{l} \times S_{l}}$ are Bernoulli variables selected from of $\{-1,1\}$ with a probability of 0.5.

\subsection{Equivariant learning}
We assume the composition of $\mathcal{A}_{\downarrow} \circ f_{\theta}$ is equivariance to group transformations. Based on the observations that PCCs are (i) rotate-invariance laterally and (ii) shift-invariance longitudinally, we subsequently apply random lateral rotations $\mathcal{G}_{r}$ and longitudinal shifts $\mathcal{G}_{s}$ to $H^{h}_{1}$. The transformed HR cube is 
\begin{equation}
	H^{h}_{2} = \mathcal{G}_{r}(\mathcal{G}_{s}(H^{h}_{1})).
\end{equation}
Then $H^{h}_{2}$ is down-sampled by $\mathcal{A}_{\downarrow}$ to obtain the transformed LR cube by $H^{l}_{2} = \mathcal{A}_{\downarrow}(H^{h}_{2})$, which is then corrupted by Possion noise by
\begin{equation}
	H^{lp}_{2} = \mathcal{P}(H^{l}_{2}/\sigma),
\end{equation}
and up-sampled by $f_\theta$ to generate the new HR output as
\begin{equation}
	H^{h}_{3} = f_{\theta}(H^{lp}_{2}). 
\end{equation}
The KL divergence-based equivariance loss is given by
\begin{equation}
	\mathcal{L}_{\mathrm{E}} =\sum_{i,j,k}^{T_{h},S_{h},S_{h}} H^{h}_{2}[i,j,k]\log \frac{H^{h}_{2}[i,j,k]}{H^{h}_{3}[i,j,k]},
	\label{Eq:e}
\end{equation}
which enforces the equivariance of $\mathcal{A}_{\downarrow} \circ f_{\theta}$. 
The total loss is defined as
\begin{equation}
	\mathcal{L}=\mathcal{L}_{\mathrm{PUKL}}+\alpha \mathcal{L}_{\mathrm{E}},
\end{equation}
where $\alpha$ is a trade-off parameter. After the training, the LR to HR propagation is $H^{h} = f_\theta(H^{l})$.
Notably, to generate a depth image $\hat{D} \in \mathbb{R}^{S_{h}\times S_{h}}$, each histogram in the HR cube is normalized by the Softmax function
\begin{equation}
	\sigma(H^{h}[i,j,k])=\frac{e^{H^{h}[i,j,k]}}{\sum_{i=1}^{T_{h}} e^{H^{h}[i,j,k]}}, 
\end{equation}
and then performed the soft Argmax function \cite{lindell2018single} by
\begin{equation}
	\hat{D}[j,k] = \sum_{i=1}^{T_{h}} iH^{h}[i,j,k].
\end{equation}
This process is equivalent to applying Maximum Likelihood Estimation (MLE) to histograms in the PCC. 

\section{Results}
\label{sec:results}
We evaluated the proposed method on the simulated dataset generated from the NYU~v2 \cite{silberman2012indoor} and real-world dataset collected by our home-built single-photon imaging system.

\begin{figure*}
	\centering
	\includegraphics[width=0.9\textwidth]{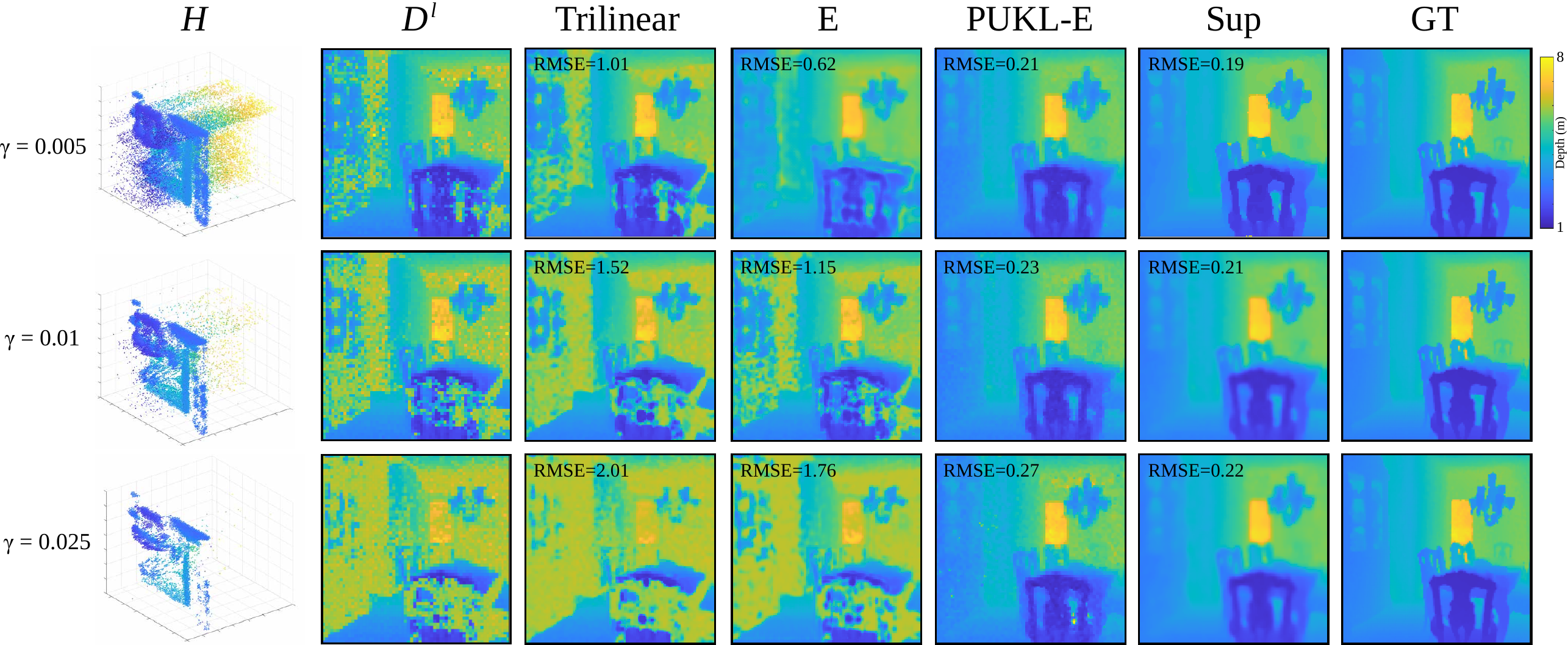}
	\caption{Results for a randomly selected scene with SBR = 1. From left to right are the original PCC, the LR depth image, the HR depth image obtained by the trilinear interpolation model, the equivariance model, the proposed model and the supervised learning model, respectively, and the ground truth. The LR depth image $D^{l}$ is obtained by MLE using the LR PCC. The RMSEs are shown in the upper left corner of the images.}
	\label{fig:r1}
	\vspace{-10pt}
\end{figure*}

\subsection{Simulated results} 
The simulated evaluation is designed for two purposes: (i) Comparison with the supervised learning. (ii) Ablation test for the PUKL and equivariance loss. Using the depth and intensity images of dining room from the NYU~v2 dataset, we generated LR data based on \eref{eq:forward}. The total number of time bins is set as $128$. The Full Width at Half Maximum (FWHM) of the emitted pulsed light covers 3 time bins. The number of training, validation and testing samples are $396$, $26$ and $104$, respectively. The Poisson noise parameter is selected from $\{0.025,0.001,0.005\}$ and the signal-to-background ratio (SBR) is selected from $\{5,1,0.2\}$. With a scale factor of $2^3$, the LR and HR cubes are with a size of $[128, 64, 64]$ and $[256, 128, 128]$, respectively. Throughout the experiment, we use the trilinear down-sampling to approximate $\mathcal{A}_{\downarrow}$ and a simplified three-dimensional version of the back-projection network in \cite{haris2018deep} to build $f_{\theta}$. 
Note that the optimal network architectures, such as multi-scale forward paths to improve the model performance, are beyond the scope of this work.

Models for comparison are the trilinear interpolation model (Trilinear), equivariance model without PUKL (E), PUKL-based model (PUKL), proposed model (PUKL-E) and supervised model based on \eref{Eq:skl} (Sup). All the models are implemented using the deep learning library Pytorch \cite{paszke2019pytorch}. We adopt the Adam optimizer with a base learning rate of $0.01$. The weight decay is $1e^{-6}$. We use $\alpha = 1$ for all experiments . Due to the GPU memory limitation, we set the batch size as 2. Each model was trained for 20 epochs on a GTX 1080ti GPU. We use the Root Mean Squared Error (RMSE) for evaluation, defined as:
\begin{equation}
	{\rm RMSE}(\hat{D}, D_{gt})=\frac{1}{S} \sqrt{\sum_{i,j}^{S}(D[i,j]-D_{gt}[i,j])^{2}}.
\end{equation}
\begin{table}[]
	\renewcommand\arraystretch{1}
	\centering
	\caption{RMSEs on the simulated dataset. Optimal and suboptimal results are highlighted in bold. }
	\label{Tab:results}
	\label{tab:lambda}
	\begin{tabular}{c|c|c|c|c|c}
		\hline
		SBR&  Trilinear & E & PUKL& PUKL-E&Sup\\ \hline
		$5$& $1.289$& $0.970$&$0.252$&$\bm{0.228}$&$\bm{0.199}$\\ \hline
		$1$ & $1.679$& $1.347$&$0.276$&$\bm{0.269}$&$\bm{0.239}$\\ \hline
		$0.2$ & $2.401$& $2.162$&$0.807$& $\bm{0.495}$&$\bm{0.479}$\\ \hline
	\end{tabular}
\end{table}

Quantitative and visualized results are presented in \tref{Tab:results} and \fref{fig:r1}, respectively. The proposed model trained by only raw measurements without labels achieves comparable performance with the supervised model with much more information. We can see that the PUKL loss greatly improves the model performance both in the RMSE and visual quality. The equivariance can provide performance gains integrated with PUKL, but cannot achieve good results on its own. As $\gamma$ increases, our model maintains a stable performance as the supervised learning model.
\begin{figure}
	\centering
	\includegraphics[width=0.45\textwidth]{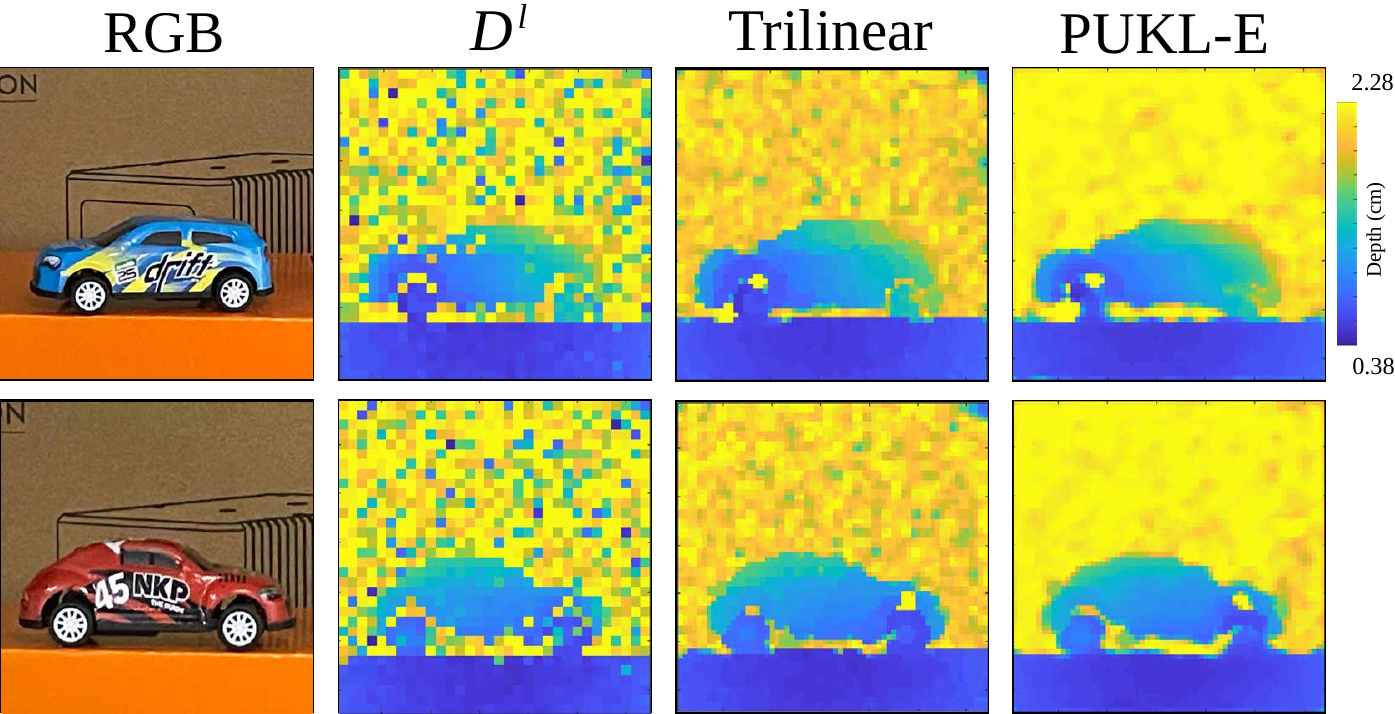}
	\caption{Testing results of real-world data. From left to right are the RGB image of the scene, the LR depth image by MLE, the HR depth image of trilinear interpolation with MLE, and the HR depth image of the proposed method.}
	\label{fig:r2}
\end{figure}

\subsection{Real-world results} 
We used the photon upconversion-based imaging system in \cite{chen2022deep} to obtain the LR raw measurements, which are not affected by the pileup effect. In the experiments, six types of car models were placed in turn about 1 meter away from the optical transceiver. The length of time bin is 1 picosecond. We measured SBR $= 0.254$ and $\gamma = 0.004$. The LR and HR cubes are with a size of $[256, 32, 32]$ and $[512, 64, 64]$, respectively. The training and testing samples are 16 and 2, respectively. As shown in \fref{fig:r2}, the proposed model generates the depth images with smoother edges and more reasonable details compared to the trilinear model. In addition, the effect of Poisson noise is greatly alleviated.

\section{Conclusion}
\label{sec:conclusion}
We have extended the robust EI framework to volumetric single-photon data using the PUKL risk estimator, which demonstrates superior SPISR performance in the presence of Poisson noise. While the method is designed to solve the SPISR problem, the idea is also promising for other inverse problems in single-photon imaging.

\bibliographystyle{IEEEtran}
\bibliography{ref}
\end{document}